\documentclass[12pt]{article}
\usepackage{amsmath}
\usepackage{amssymb} 
\usepackage[small]{caption2} 
\usepackage{fleqn} 
\usepackage{graphicx} 
\usepackage{mathrsfs}	
\usepackage[small,loose]{subfigure}  
\usepackage{cite} 
\advance \headheight by 3.0truept

\newcommand{\CenterObject}[1]{\ensuremath{\vcenter{\hbox{#1}}}}
\newcommand{\D}{\mathrm{d}}
\newcommand{\I}{\mathrm{i}}
\newcommand{\stau}{\widetilde{\tau}}

\begin{document}
\title{{\normalsize DESY 04-043\hfill\mbox{}\\
\normalsize UT 04-11\hfill\mbox{}\\
March 2004\hfill\mbox{}}\\
\vspace{1cm}
\textbf{Gravitino and Goldstino at Colliders}\footnote{Contribution to
the LHC / LC study group report, eds.\ G.~Weiglein et al.}\\[8mm]}
\author{Wilfried Buchm\"uller$^{(a)}$, Koichi Hamaguchi$^{(a)}$, \\
Michael Ratz$^{(a)}$ and Tsutomu Yanagida$^{(b)(c)}$ \\[0.4cm]
{\normalsize\textit{$^{(a)}$Deutsches Elektronen-Synchrotron DESY, 
22603 Hamburg, Germany}}\\
{\normalsize\textit{$^{(b)}$Department of Physics, University of Tokyo, 
Tokyo 113-0033, Japan}}\\
{\normalsize\textit{$^{(c)}$Research Center for the Early Universe,  
University of Tokyo,
Japan}}
}
\date{}
\maketitle
\thispagestyle{empty}

\begin{abstract}
\noindent
We consider theories with spontaneously broken global or local
supersymmetry where the pseudo-goldstino or the gravitino is the
lightest superparticle (LSP).  Assuming that the long-lived
next-to-lightest superparticle (NSP) is a charged slepton, we study
several supergravity predictions: the NSP lifetime, angular and energy
distributions in 3-body NSP decays. The characteristic couplings of
the gravitino, or goldstino, can be tested even for very small masses.
\end{abstract}

\newpage
\subsection*{Introduction}

The discovery of supersymmetry at the Tevatron, the LHC or a future
Linear Collider would raise the question how supersymmetry is realized
in nature. Clearly, supersymmetry is broken. Spontaneously broken
global supersymmetry would predict the existence of a spin-1/2
goldstino ($\chi$) whereas the theoretically favoured case of local
supersymmetry requires a massive spin-3/2 gravitino ($\psi_{3/2}$).

In a recent paper \cite{BHRY} we have studied how a massive gravitino,
if it is the lightest superparticle (LSP), may be discovered in decays
of $\stau$, the scalar $\tau$ lepton, which is naturally the
next-to-lightest superparticle (NSP). The determination of gravitino
mass and spin appears feasible for gravitino masses in the range from
about $1\,\mathrm{GeV}$ to $100\,\mathrm{GeV}$.  As we shall discuss
in this note, evidence for the characteristic couplings of a
pseudo-goldstino, which corresponds to the spin-1/2 part of the gravitino, 
can be obtained even for masses much smaller than $1\,\mathrm{GeV}$.  

The gravitino mass depends on the mechanism of supersymmetry breaking.
It can be of the same order as other superparticle masses, like in
gaugino mediation~\cite{Kaplan:1999ac} or gravity
mediation~\cite{Nilles:1984ge}. But it might also be much smaller as
in gauge mediation scenarios \cite{Giudice:1998bp}. As LSP, the
gravitino is an attractive dark matter candidate~\cite{Bolz:1998ek}.

The $\stau$ NSP has generally a long lifetime because of the small,
Planck scale suppressed coupling to the gravitino LSP.  The production
of charged long-lived heavy particles at colliders is an exciting
possibility~\cite{Drees:1990yw}.  They can be directly produced in
pairs or in cascade decays of heavier superparticles.  In the context
of models with gauge mediated supersymmetry breaking the production of
slepton NSPs has previously studied for the
Tevatron~\cite{Feng:1997zr}, for the LHC~\cite{Ambrosanio:2000ik} and
for a Linear Collider~\cite{Ambrosanio:1999iu}.

The dominant $\stau$ NSP decay channel is $\stau\to\tau+\text{missing
energy}$.  In the following we shall study how to identify the
gravitino or goldstino as carrier of the missing energy. First, one
will measure the NSP lifetime. Since the gravitino couplings are fixed
by symmetry, the NSP lifetime is predicted by supergravity given the
gravitino mass, which can be inferred from kinematics. Second, one can
make use of the 3-body NSP decay $\stau\to\tau+\gamma+X$ where
$X=\psi_{3/2}$ or $X=\chi$.  The angular and energy distributions and
the polarizations of the final state photon and lepton carry the
information on the spin and couplings of gravitino or goldstino.

For gravitino masses in the range from about $10\,\mathrm{keV}$ to
$100\,\mathrm{GeV}$, the NSP is essentially stable for collider
experiments, and one has to accumulate the NSPs to study their
decay. Sufficiently slow, strongly ionizing sleptons will be stopped
within the detector. One may also be able to collect faster sleptons
in a storage ring.  For gravitino masses less than
$\mathcal{O}(10\,\mathrm{keV})$ the $\stau$ can decay inside the
detector, which may be advantageous from the experimental point of
view.

At LHC one expects $\mathcal{O}(10^6)$ NSPs per year which are mainly
produced in cascade decays of squarks and gluinos
\cite{Beenakker:1997ch}.  The NSPs are mostly produced in the forward
direction~\cite{Maki:1998ih} which should make it easier to accumulate
$\stau$s in a storage ring.  In a Linear Collider an integrated
luminosity of $500\,\mathrm{fb}^{-1}$ will yield $\mathcal{O}(10^5)$
$\stau$s
\cite{Aguilar-Saavedra:2001rg}. Note that, in a Linear Collider, one
can also tune the velocity of the produced $\stau$s by adjusting the
$e^+e^-$ center-of-mass energy. A detailed study of the possibilities
to accumulate $\stau$ NSPs is beyond the scope of this note. In the
following we shall assume that a sufficiently large number of $\stau$s
can be produced and collected.

This study is strongly based on Ref.~\cite{BHRY}. Here, we discuss
in more detail the case of a very light gravitino, or
pseudo-goldstino, for which the $\stau$ NSP can decay inside the
detector.  Although in this case it is difficult to determine mass and
spin of the gravitino, one can still see the characteristic coupling
of the gravitino, which is essentially the goldstino coupling, via the
3-body decay $\widetilde{\tau}\to\tau+\gamma+X$ with $X=\psi_{3/2}$ or
$X=\chi$.

\subsection*{Planck mass from $\boldsymbol{\stau}$ decays}

The $\stau$ decay rate is dominated by the two-body decay into $\tau$
and gravitino,
\begin{eqnarray}
 \Gamma_{\stau}^\mathrm{2-body}
 & = &
 \frac{\left( m_{\stau}^2 - m_{3/2}^2 - m_{\tau}^2 \right)^4 }{
	48\pi\,m_{3/2}^2\,M_\mathrm{P}^2\,m_{\stau}^3 }\,
 \left[1-\frac{4m_{3/2}^2\,m_{\tau}^2}{
 	\left( m_{\stau}^2 - m_{3/2}^2 - m_{\tau}^2 \right)^2} \right]^{3/2}\;,
 \label{eq:2bodyDecayRateWithTau}
\end{eqnarray}
where $M_\mathrm{P}=(8\pi\, G_\mathrm{N})^{-1/2}$ denotes the reduced
Planck mass, $m_\tau=1.78\,\mathrm{GeV}$ is the $\tau$ mass,
$m_{\stau}$ is the $\stau$ mass, and $m_{3/2}$ is the gravitino
mass. For instance, $m_{\stau}=150\,\mathrm{GeV}$ and
$m_{3/2}=10\,\mathrm{keV}$ leads to a lifetime of
$\Gamma_{\stau}^{-1}\simeq 7.8\times 10^{-7}\,\mathrm{s}$, and
$m_{\stau}=150\,\mathrm{GeV}$ and $m_{3/2}=75\,\mathrm{GeV}$ results
in $\Gamma_{\stau}^{-1}\simeq 4.4\,\mathrm{y}$.

Since the decay rate depends only on two unknown masses $m_{\stau}$
and $m_{3/2}$, independently of other SUSY parameters, gauge and
Yukawa couplings, it is possible to test the prediction of the
supergravity if one can measure these masses. The mass $m_{\stau}$ of
the NSP will be measured in the process of accumulation.  Although the
outgoing gravitino is not directly measurable, its mass can also be
inferred kinematically unless it is too small,
\begin{eqnarray}
  m_{3/2}^2 &=& m_{\stau}^2 + m_\tau^2 -2m_{\stau} E_\tau\;.
\end{eqnarray}
The gravitino mass can be determined with the same accuracy as
$E_\tau$ and $m_{\stau}$, i.e.\ with an uncertainty of a few GeV.

Once the masses $m_{\stau}$ and $m_{3/2}$ are measured, one can
compare the predicted decay rate \eqref{eq:2bodyDecayRateWithTau} with
the observed decay rate, thereby testing an important supergravity
prediction.  In other words, one can determine the `supergravity
Planck scale' from the NSP decay rate which yields, up to
$\mathcal{O}(\alpha)$ corrections,
\begin{eqnarray}
 M_\mathrm{P}^2(\text{supergravity}) & = &
 \frac{\left( m_{\stau}^2 - m_{3/2}^2 - m_{\tau}^2 \right)^4 }{
	48\pi\,m_{3/2}^2\,m_{\stau}^3\, \Gamma_{\stau}}\,
 \left[1-\frac{4m_{3/2}^2\,m_{\tau}^2}{
 	\left( m_{\stau}^2 - m_{3/2}^2 - m_{\tau}^2 \right)^2} \right]^{3/2}\;.
\end{eqnarray}
The result can be compared with the Planck scale of Einstein gravity,
i.e.\ Newton's constant determined by macroscopic measurements,
$G_\mathrm{N}=6.707(10)\cdot10^{-39}\,\mathrm{GeV}^{-2}$
\cite{Hagiwara:2002fs},
\begin{eqnarray}
  M_\mathrm{P}^2(\mathrm{gravity}) &=& (8\pi\, G_\mathrm{N})^{-1} 
  \,=\, (2.436(2)\cdot 10^{18}\,\mathrm{GeV})^2\;.
\end{eqnarray}
The consistency of the microscopic and macroscopic determinations of
the Planck scale is an unequivocal test of supergravity.

Note that the measurement of the gravitino mass yields another
important quantity in supergravity, the mass scale of spontaneous
supersymmetry breaking $M_\mathrm{SUSY}\,=\,
\sqrt{\sqrt{3}M_\mathrm{P}\,m_{3/2}}$.  This is the analogue of the
Higgs vacuum expectation value $v$ in the electroweak theory, where $v
= \sqrt{2}m_W/g = (2\sqrt{2}G_\mathrm{F})^{-1/2}$.

\subsection*{Gravitino and goldstino versus neutralino}

If the measured decay rate and the kinematically determined mass of
the invisible particle are consistent with
Eq.~\eqref{eq:2bodyDecayRateWithTau}, one already has strong evidence
for supergravity and the gravitino LSP. To analyze the couplings of
the invisible particle, one can study the 3-body decay
$\stau\to\tau+\gamma+X$ for the gravitino $X=\psi_{3/2}$ and compare
it with the case where $X$ is a hypothetical spin-1/2 neutralino.
This is of particular importance if the mass of the invisible particle
is so small that the supergravity prediction for the NSP lifetime, 
as described in the previous section, cannot be tested.

The NSP $\stau$ is in general a linear combination of
$\stau_\mathrm{R}$ and $\stau_\mathrm{L}$, the superpartners of the
right-handed and left-handed $\tau$ leptons $\tau_\mathrm{R}$ and
$\tau_\mathrm{L}$, respectively.  The interaction of the gravitino
$\psi_{3/2}$ with scalar and fermionic $\tau$ leptons is described by
the lagrangian \cite{Wess:1992cp},
\begin{eqnarray}
 \mathscr{L}_{3/2}
 & = & 
 -\frac{1}{\sqrt{2}M_\mathrm{P}}
 \left[
   (D_\nu\,\stau_\mathrm{R})^*\overline{\psi^\mu}\,\gamma^\nu\,\gamma_\mu\,
   P_\mathrm{R} \tau
   +
   (D_\nu\,\stau_\mathrm{R})\,\overline{\tau} P_\mathrm{L}
   \gamma_\mu\,\gamma^\nu\,\psi^\mu\right]\;,
 \label{eq:FullGravitinoLagrangian}
\end{eqnarray}
where $D_\nu\,\stau_\mathrm{R} = (\partial_\nu + \I e\, A_\nu)
\stau_\mathrm{R}$ and $A_\nu$ denotes the gauge boson.  The
interaction lagrangian of $\stau_\mathrm{L}$ has an analogous form.

As an example for the coupling of a hypothetical spin-1/2 neutralino
to $\stau$ and $\tau$, we consider the Yukawa
interaction\footnote{This interaction would arise from gauging the
anomaly free U(1) symmetry $L_{\tau}-L_{\mu}$, the difference of
$\tau$- and $\mu$-number, in the MSSM, with $\lambda$ being the
gaugino.},
\begin{equation}
 \mathscr{L}_\mathrm{Yukawa}\,=\,
  h\left(\stau_\mathrm{R}^*\,\overline{\lambda}\,P_\mathrm{R}\,\tau
         +\stau_\mathrm{L}^*\,\overline{\lambda}\,P_\mathrm{L}\,\tau \right)
+\text{h.c.}\;.
\label{eq:LYukawa}
\end{equation}
Note that for very small coupling $h$, the $\stau$ decay rate could
accidentally be consistent with the supergravity prediction
Eq.~(\ref{eq:2bodyDecayRateWithTau}).

Also the goldstino $\chi$ has Yukawa couplings of the type given in
Eq.~(\ref{eq:LYukawa}).  The full interaction lagrangian is obtained
by performing the substitution $\psi_\mu\to \sqrt{{2\over 3}}{1\over
m_{3/2}}\partial_\mu\chi$ in the supergravity lagrangian. The 
non-derivative form of the effective lagrangian for $\chi$ is given by
\cite{Lee:1998aw},
\begin{equation}
 \mathscr{L}_\mathrm{eff}
 \,=\,
 \frac{m_{\widetilde{\tau}}^2}{\sqrt{3}M_\mathrm{P}\,m_{3/2}}
 \left(\widetilde{\tau}_\mathrm{R}^*\,\overline{\chi}\,P_\mathrm{R}\,\tau
 +\widetilde{\tau}_\mathrm{R}\,\overline{\tau}\,P_\mathrm{L}\,\chi\right)
 -\frac{m_{\widetilde{\gamma}}}{4\sqrt{6}M_\mathrm{P}\,m_{3/2}}
 \overline{\chi}[\gamma^\mu,\gamma^\nu]\,\widetilde{\gamma}\,F_{\mu\nu}\;,
 \label{eq:GoldstinoLagrangian}
\end{equation}
where we have neglected a quartic interaction term which is irrelevant
for our discussion. In the following, we consider a massive
pseudo-goldstino $\chi$, in order to compare it with the massive
gravitino and neutralino.  Like a pseudo-Goldstone boson, the
pseudo-goldstino has goldstino couplings and a mass which explicitly
breaks global supersymmetry.

\begin{figure}[t]
\begin{center}
 \subfigure[{}]{
 \CenterObject{\includegraphics[scale=1]{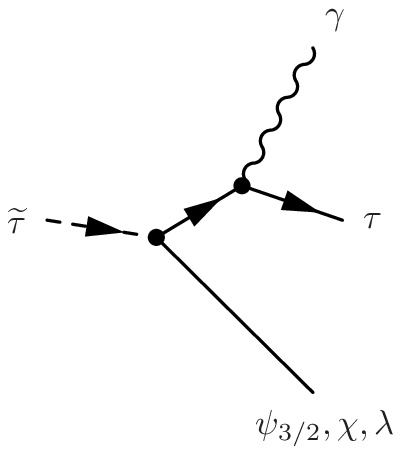}}}
\hfil
 \subfigure[{}]{
 \CenterObject{\includegraphics[scale=1]{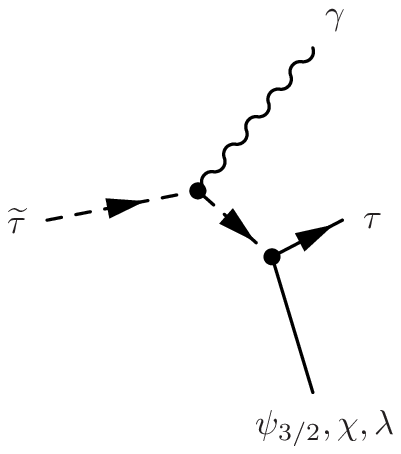}}}
 \\
 \subfigure[{}\label{fig:PhotinoContribution}]{
 \CenterObject{\includegraphics[scale=1]{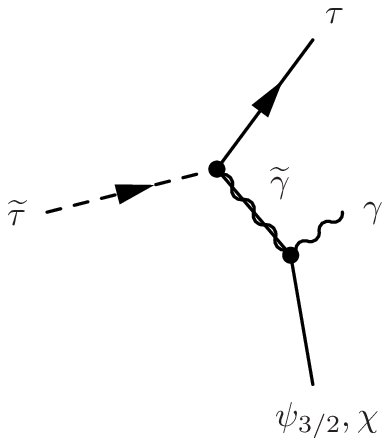}}}
 \hfil
 \subfigure[{}\label{fig:4piontVertex}]{
 \CenterObject{\includegraphics[scale=1]{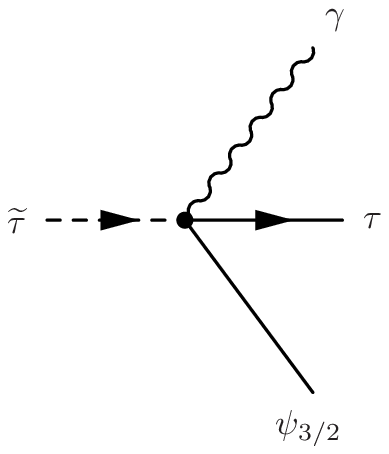}}}
\end{center} 
 \caption{Diagrams contributing to the 3-body decay
 $\widetilde{\tau}\to\tau+ \gamma+X$ where
 $X=\psi_{3/2},\chi,\lambda$. In the limit of very large
 $m_{\widetilde{\gamma}}$, diagram (c) becomes irrelevant for the
 gravitino, but it always contributes for the goldstino where it leads
 effectively to a 4-point interaction.}
 \label{fig:3bodystauR2tauRgammaGoldstino}
\end{figure}

Note that the goldstino coupling to the photon supermultiplet is
proportional to the photino mass $m_{\widetilde{\gamma}}$. As a
consequence, the contribution to 3-body $\stau$-decay with
intermediate photino (cf.\ Fig.~\ref{fig:PhotinoContribution}) is not
suppressed for very large photino masses. As we shall see, this leads
to significant differences between the angular distributions for pure
Yukawa and goldstino couplings, even when $\chi$ and $\lambda$ are
very light.

\begin{figure}[h!]
\begin{center}
 \subfigure[Kinematical configuration. The arrows denote the momenta.
 \label{fig:KinematicalConfiguration}]{
 	\CenterObject{\includegraphics{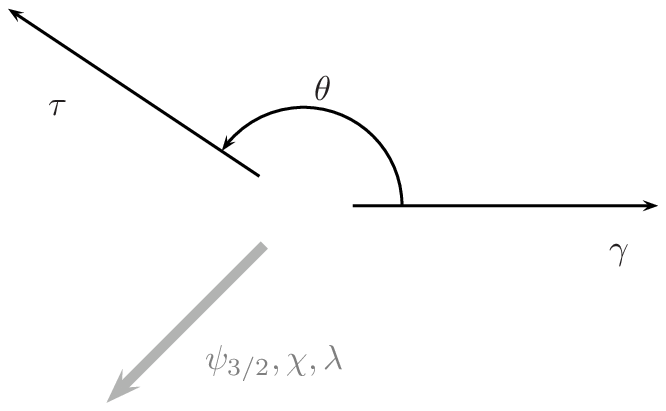}}}
 \quad
 \subfigure[Gravitino $\psi_{3/2}$]{
 	\CenterObject{\includegraphics[scale=0.85]{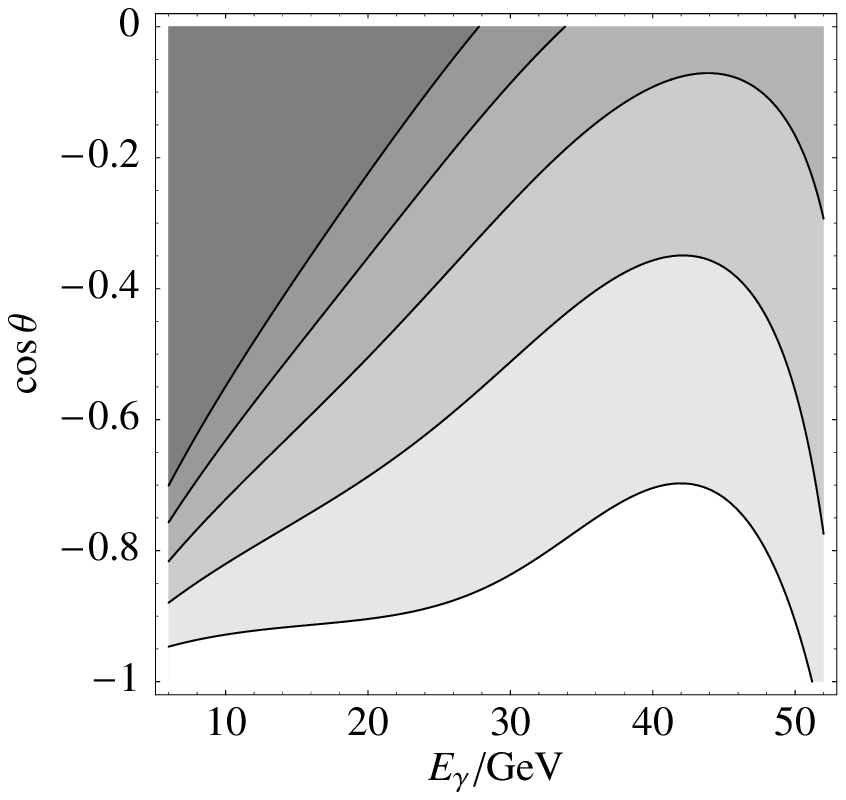}}}
 \\
 \subfigure[Pseudo-goldstino $\chi$]{
 	\CenterObject{\includegraphics[scale=0.85]{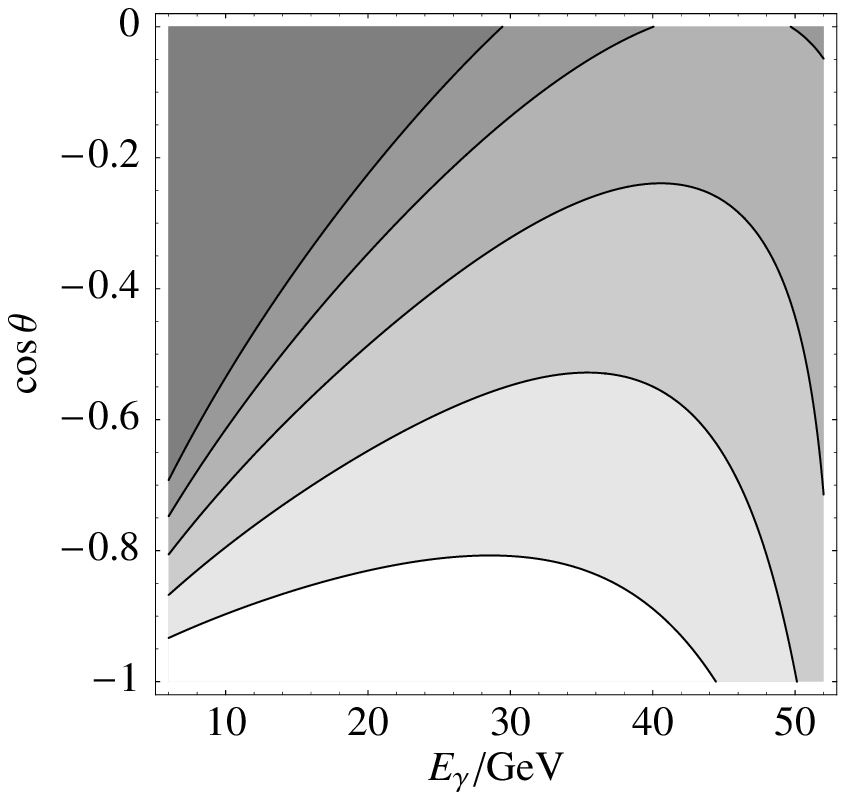}}}
 \quad
 \subfigure[Spin-1/2 neutralino $\lambda$]{
 	\CenterObject{\includegraphics[scale=0.85]{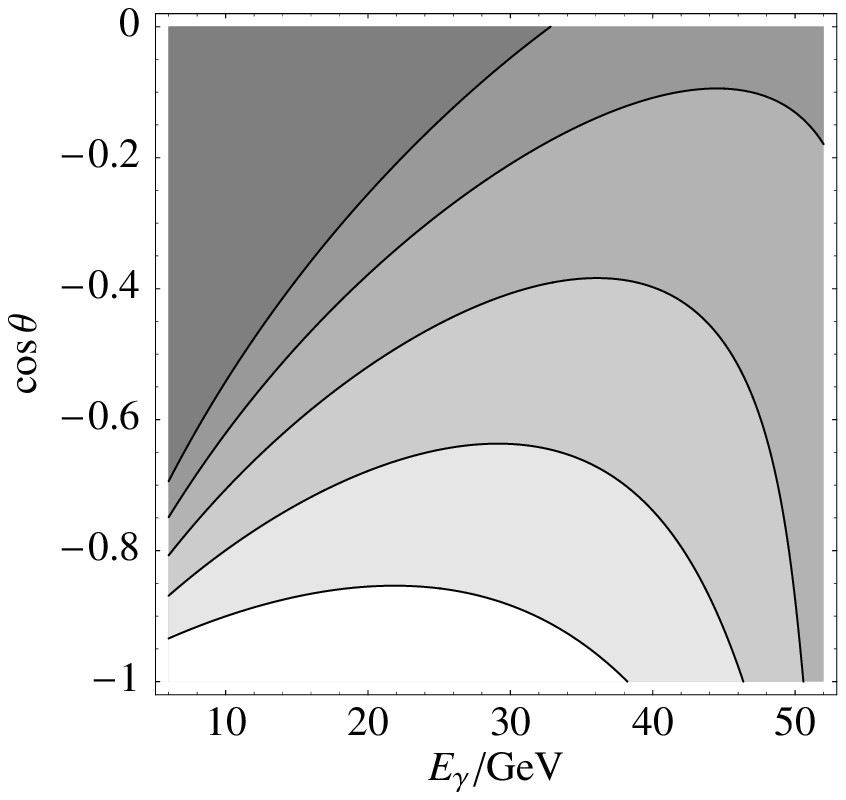}}}
\end{center}
\caption{(a) shows the kinematical configuration of the 3-body decay.
  The others are contour plots of the differential decay rates for (b)
  gravitino $\psi_{3/2}$, (c) pseudo-goldstino $\chi$ and (d)
  neutralino $\lambda$.  $m_{\widetilde{\tau}}=150\,\mathrm{GeV}$ and
  $m_{X}=75\,\mathrm{GeV}$ ($X=\psi_{3/2},\lambda,\chi$).  The
  boundaries of the different gray shaded regions (from bottom to top)
  correspond to $\Delta(E_\gamma ,\cos\theta)[\mathrm{GeV}^{-1}]
  =10^{-3}, 2\times10^{-3}, 3\times10^{-3}, 4\times10^{-3},
  5\times10^{-3}$.  Darker shading implies larger rate.}
\label{fig:CompareDifferentialDecayRate}
\end{figure}

In $\stau$ decays both, photon and $\tau$ lepton will mostly be very
energetic.  Hence the photon energy $E_{\gamma}$ and the angle
$\theta$ between $\tau$ and $\gamma$ can be well measured (cf.\
Fig.~\ref{fig:KinematicalConfiguration}).  We can then compare the
differential decay rate
\begin{equation}
 \Delta(E_\gamma ,\cos\theta)\,=\,{1\over\alpha\,\Gamma_{\widetilde{\tau}}}
 \frac{\D^2\Gamma(\widetilde{\tau}\to\tau+\gamma+X)}{\D E_\gamma\,\D \cos\theta}\;,
\end{equation}
for the gravitino LSP ($X=\psi_{3/2}$), the pseudo-goldstino
($X=\chi$) and the hypothetical neutralino ($X=\lambda$). 
Details of the calculation are given in
Ref.~\cite{BHRY}.  The differences between $\psi_{3/2}$, $\chi$ and
$\lambda$ become significant in the backward direction ($\cos\theta <
0$) as demonstrated by Fig.~\ref{fig:CompareDifferentialDecayRate}
(b)-(d), where $m_{\widetilde{\tau}}=150\,\mathrm{GeV}$ and
$m_{X}=75\,\mathrm{GeV}$ ($X=\psi_{3/2},\chi,\lambda$). The three
differential distributions are qualitatively different and should
allow to distinguish experimentally gravitino, goldstino and
neutralino.

Let us now consider the case of small $m_X$. Then  the goldstino lagrangian
\eqref{eq:GoldstinoLagrangian} effectively describes the
gravitino interactions. Therefore, one can no longer distinguish
between gravitino and goldstino in this case.  However, even for small
$m_X$ one can discriminate the gravitino or goldstino from the
neutralino. The difference between goldstino $\chi$ and neutralino
$\lambda$ stems from the photino contribution (cf.\
Fig.~\ref{fig:PhotinoContribution}) which does not decouple for large
photino mass $m_{\widetilde{\gamma}}$.  This is different from the
gravitino case where the analogous diagram becomes irrelevant in the
limit $m_{\widetilde{\gamma}}\gg m_{\stau}$ which is employed
throughout this study. 

\begin{figure}[t]
 \begin{center}
  \subfigure[Gravitino/goldstino]{%
    \CenterObject{\includegraphics[scale=0.85]{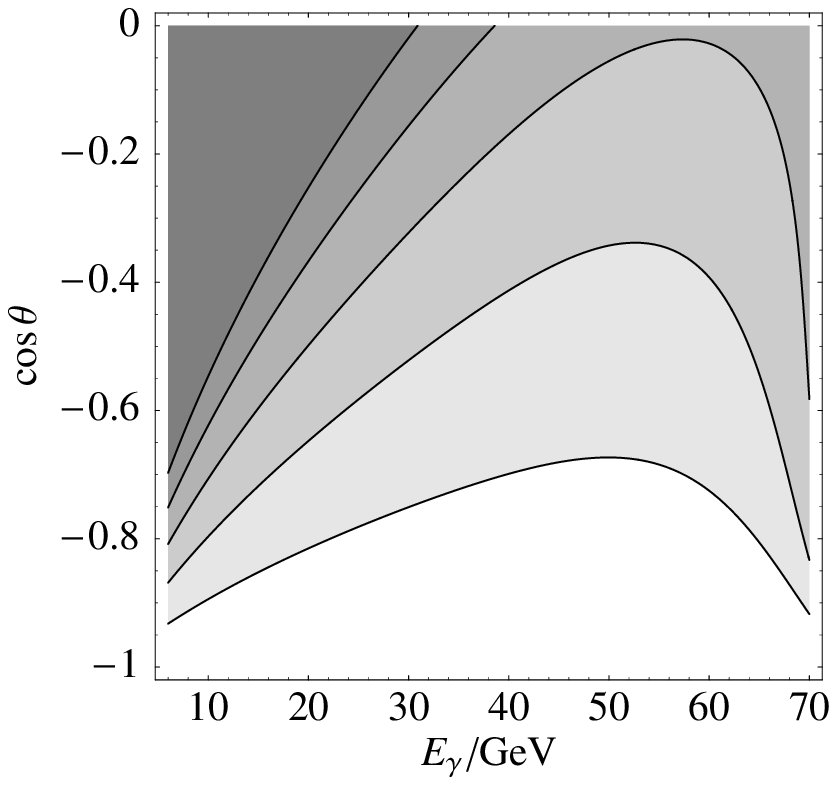}}
	}
	\hfil
   \subfigure[Neutralino]{%
    \CenterObject{\includegraphics[scale=0.85]{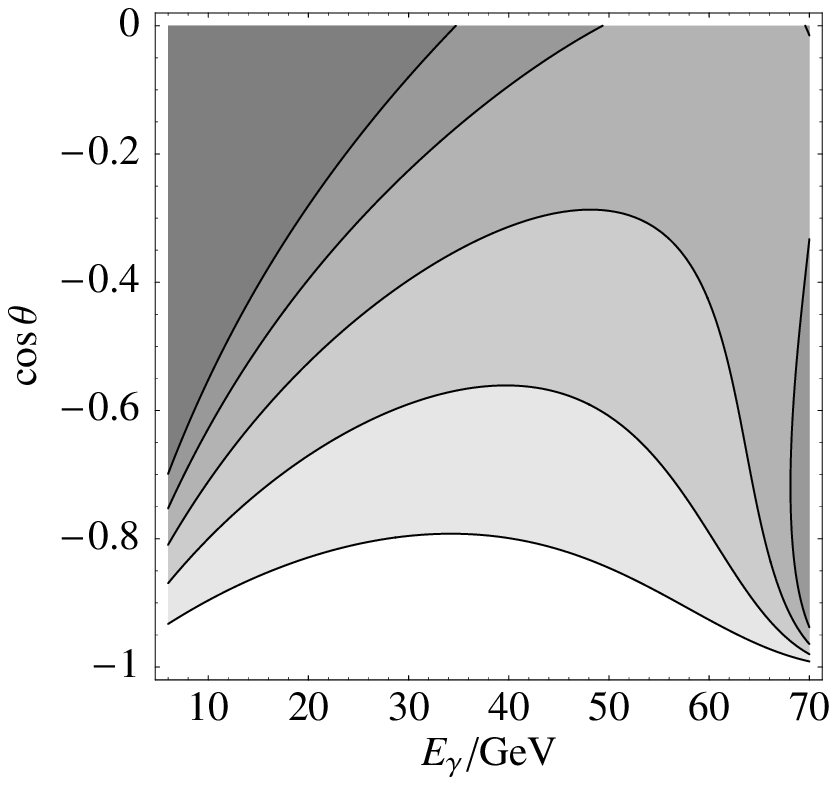}}
	}
  \end{center}
  \caption{Contour plots of the differential decay rates for (a)
  gravitino $\psi_{3/2}$ and (b) neutralino $\lambda$.
  $m_{\widetilde{\tau}}=150\,\mathrm{GeV}$, $m_{X}= 0.1~\mathrm{GeV}$
  ($X=\psi_{3/2},\lambda$). The figures remain essentially the
  same as long as $r=m_X^2/m_{\stau}^2\ll 1$. The contours have the same
  meaning as in Fig.~\ref{fig:CompareDifferentialDecayRate}.}
  \label{fig:DiffDecayRateGoldstino}
\end{figure}

The arising discrepancy between gravitino or goldstino and neutralino
is demonstrated by Fig.~\ref{fig:DiffDecayRateGoldstino}. It clearly
shows that even for very small masses $m_{3/2}$ and $m_\lambda$, the
differential decay rates $\Delta$ for gravitino $\psi_{3/2}$ and
neutralino $\lambda$ are distinguishable. This makes it possible to
discriminate gravitino and goldstino from a hypothetical neutralino
even for very small masses. Note that the plots of 
Fig.~\ref{fig:DiffDecayRateGoldstino} remain essentially the same as long as 
$r=m_X^2/m_{\stau}^2\ll 1$.

\clearpage

Let us finally comment on the experimental feasibility to determine
gravitino or goldstino couplings. The angular distribution of the
3-body decay is peaked in forward direction ($\theta=0$).  Compared to
the 2-body decay, backward ($\cos\theta<0$) 3-body decays are
suppressed by $\sim 10^{-1}\times\alpha\simeq10^{-3}$. Requiring
10\dots100 events for a signal one therefore needs $10^4$ to $10^5$
$\stau$s, which appears possible at the LHC and also at a Linear
Collider according to the above discussion.

\subsection*{Gravitino spin}

\begin{figure}[!ht]
\begin{center}
 \subfigure[Characteristic spin-$3/2$ process. The thick arrows represent the
 	spins.\label{fig:TypicalSpin32}]{
 	\CenterObject{\includegraphics{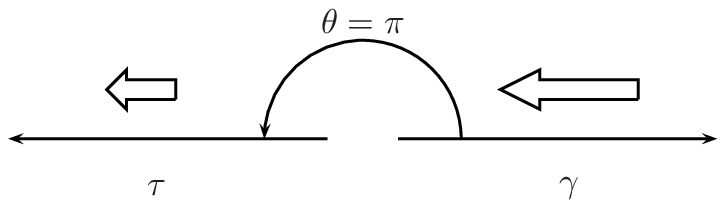}}}
 \hfil
 \subfigure[$m_X=10\,\mathrm{GeV}$.]{
 	\CenterObject{\includegraphics[scale=0.85]{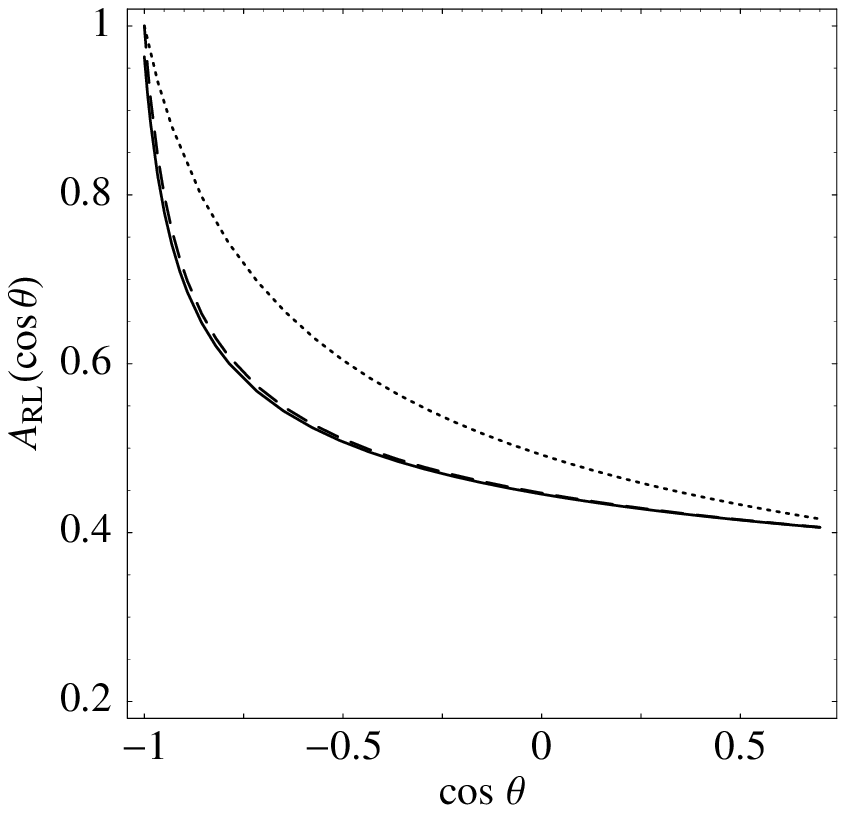}}}
 \\
 \subfigure[$m_X=30\,\mathrm{GeV}$.]{
 	\CenterObject{\includegraphics[scale=0.85]{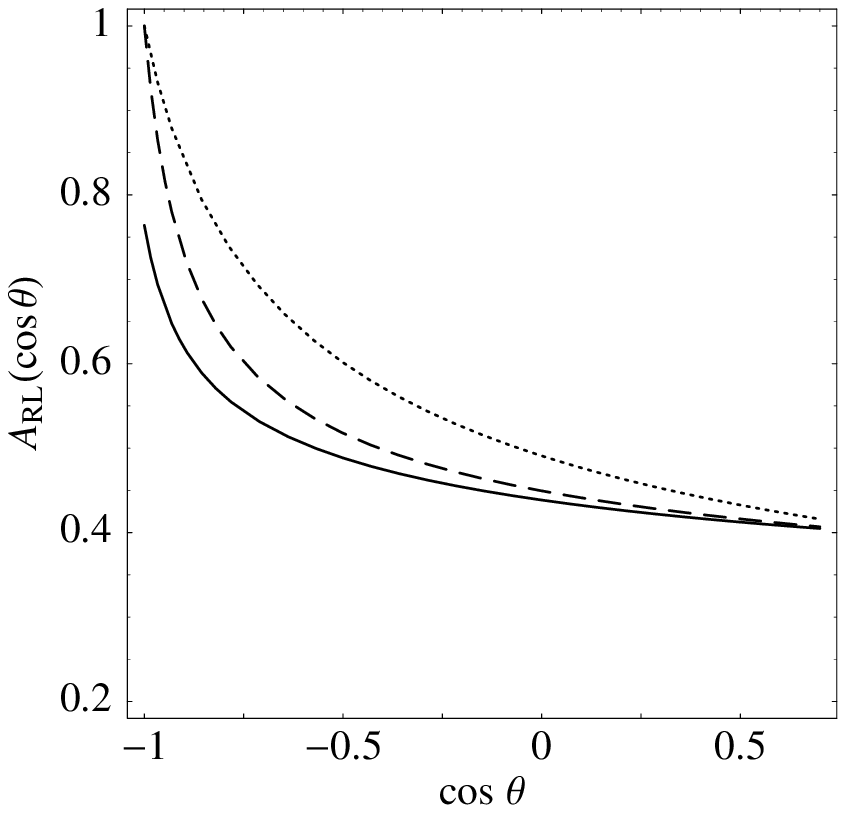}}}
 \hfil
 \subfigure[$m_X=75\,\mathrm{GeV}$.]{
 	\CenterObject{\includegraphics[scale=0.85]{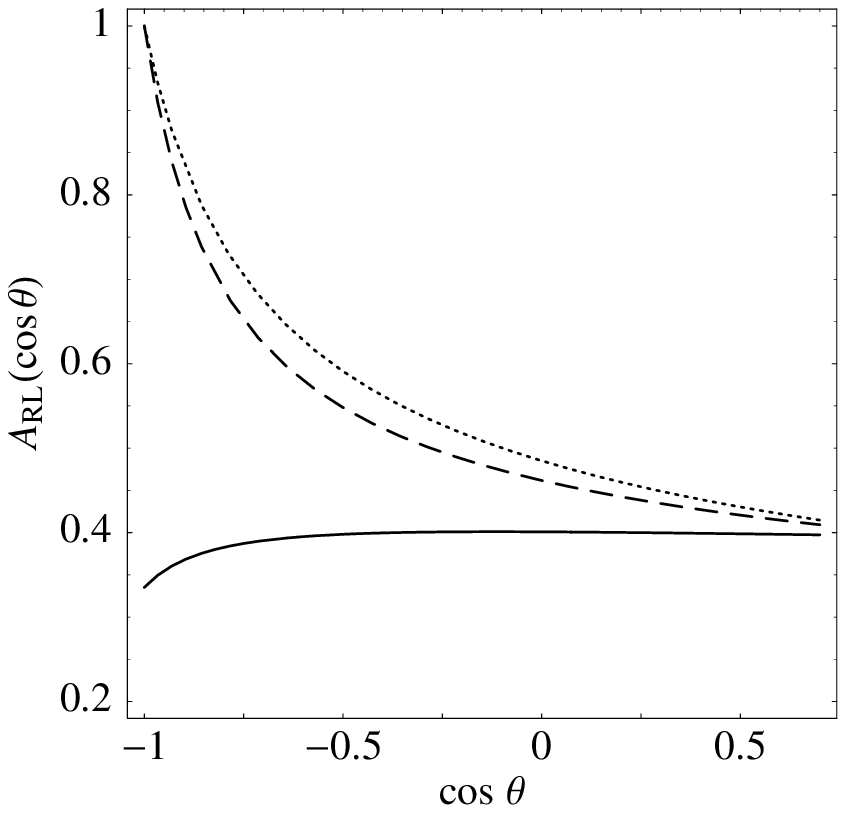}}}
\end{center} 
\caption{(a) illustrates the characteristic spin-3/2 process: 
 photon and $\tau$ lepton move in opposite directions and the spins
 add up to $3/2$, so the invisible particle also has spin $3/2$.  The
 other figures show angular asymmetries for gravitino $\psi_{3/2}$
 (solid curve), goldstino $\chi$ (dashed curve) and neutralino
 $\lambda$ (dotted curve).  $m_{\stau}=150\,\mathrm{GeV}$. The photon
 energy is larger than $10\%$ of the maximal kinematically allowed
 energy (cf.\ Ref.~\cite{BHRY}).  Note that the asymmetries only
 depend on the ratio $r=m_X^2/m_{\stau}^2$ ($X=\psi_{3/2},
 \chi,\lambda$).}
\label{fig:CompareLandR}
\end{figure}

A third test of supergravity is intuitively more straightforward
though experimentally even more challenging than the previous ones. It
is again based on 3-body decays. We now take into account also the
polarizations of the visible particles, $\gamma$ and $\tau$. The main
point is obvious from Fig.~\ref{fig:TypicalSpin32} where a left-handed
photon and a right-handed $\tau$ move in opposite
directions.\footnote{For simplicity, we here restrict ourselves to the
case of a right-handed $\stau$ LSP, leaving finite left-right mixing
angles for future investigations.} Clearly, this configuration is
allowed for an invisible spin-3/2 gravitino but it is forbidden for a
spin-1/2 goldstino or neutralino.  Unfortunately, measuring the
polarizations is a difficult task.

\clearpage
As Fig.~\ref{fig:TypicalSpin32} illustrates, the spin of the invisible
particle influences the angular distribution of final states with
polarized photons and $\tau$ leptons.  An appropriate observable is
the angular asymmetry
\begin{equation}
 A_\mathrm{RL}(\cos\theta)\,=\,
 \frac{\displaystyle 
 \frac{\D \Gamma}{\D\cos\theta}(\stau_\mathrm{R}
 \to\tau_\mathrm{R}+\gamma_\mathrm{R}+X)
 -\frac{\D \Gamma}{\D\cos\theta}(\stau_\mathrm{R}
 \to\tau_\mathrm{R}+\gamma_\mathrm{L}+X)}{\displaystyle
 \frac{\D \Gamma}{\D\cos\theta}(\stau_\mathrm{R}
 \to\tau_\mathrm{R}+\gamma_\mathrm{R}+X)
 +\frac{\D \Gamma}{\D\cos\theta}(\stau_\mathrm{L}
 \to\tau_\mathrm{R}+\gamma_\mathrm{L}+X)}\;,
\end{equation}
where $X$ denotes gravitino ($X = \psi_{3/2}$), goldstino ($X=\chi$)
or neutralino ($X=\lambda$). Note that, as discussed before, the
photino does not decouple in the case $X=\chi$.

The three angular asymmetries are shown in Fig.~\ref{fig:CompareLandR}
for $m_{\stau}=150\,\mathrm{GeV}$ and different masses of the
invisible particle.  As expected, the decay into right-handed $\tau$
and left-handed photon at $\theta=\pi$ is forbidden for spin-1/2
invisible particles ($\chi$ and $\lambda$), whereas it is allowed for
the spin-3/2 gravitino. This is clearly visible in
Figs.~\ref{fig:CompareLandR}(c) and (d); for small gravitino masses
the goldstino component dominates the gravitino interaction as
illustrated by Fig.~\ref{fig:CompareLandR}(b). The discrepancy between
gravitino and goldstino compared to a hypothetical neutralino persists
for arbitrarily small $m_X$, which is analogous to the double
differential distribution discussed in the previous section.

\subsection*{Conclusions}

We have discussed how one may discover a massive gravitino, and
thereby supergravity, at the LHC or a future Linear Collider, if the
gravitino is the LSP and a charged slepton is the NSP. With the
gravitino mass inferred from kinematics, the measurement of the NSP
lifetime will test an unequivocal prediction of supergravity. The
analysis of 3-body NSP decays will reveal the couplings of the
gravitino or the goldstino. For very small masses, one can distinguish
the gravitino from the neutralino but not from the goldstino.  For
masses larger than about $1\,\mathrm{GeV}$, the determination of 
gravitino mass and spin appears feasible.
 
\subsection*{Acknowledgements}

We would like to thank T.~Plehn, P.~Schleper 
and P.~M.~Zerwas for valuable discussions.

\end{document}